\input{aipcheck}

\documentclass[
    ,final            % use final for the camera ready runs
  ,numberedheadings % uncomment this option for numbered sections
  ]
  {aipproc}

\layoutstyle{8x11single}

\newcommand{\be}{\begin{equation}}
\newcommand{\ee}{\end{equation}}

\newcommand{\beq}{\begin{eqnarray}}
\newcommand{\eeq}{\end{eqnarray}}

\begin{document}

\title{Cosmic Superstrings: Dynamics and Cusps}


\author{William Nelson\footnote{This work was done in collaboration with Anne-Christine Davis, Senthooran Rajamanoharan, and Mairi
Sakellariadou~\cite{Davis:2008kg}.} }{address={ King's College London, Strand, London, WC2R 2LS, U.\ K.\  \\
Institute of Gravitation and the Cosmos, Penn State University, State College, PA 16801, U.\ S.\ A.\ } }

\begin{abstract}
Whilst standard field theoretic Cosmic Strings cannot end, Cosmic Superstrings can form three string junctions,
at which each string ends. This opens up a new class of possible boundary conditions for such strings and 
we show that, at least when the junctions are close together, a string ending of two such junctions will
generically have cusps. Cusps are of particular interest as they are strong emitters of radiation (both
gravitational and particle) and hence are possible observables. The detection of cusps from Cosmic Superstrings
between junctions would be a rare observational window into the realm of String theory and Brane inflation models.
\end{abstract}

\maketitle

%%%%%%%%%%%%%%%%%%%%%%%%%%%%%%%%%%%%%%%%%%%%
%% MAINMATTER
%%%%%%%%%%%%%%%%%%%%%%%%%%%%%%%%%%%%%%%%%%%%

\section{Introduction}\label{sec:intro}

Standard\footnote{i.e.\ Albelian Cosmic Strings, Non-Albelian versions can also have junctions and are a useful
way of modeling the dynamics of Cosmic Superstrings. See Ref.~\cite{Bevis:2008hg} for some recent work in this area.} 
field theoretic Cosmic Strings form during symmetry breaking phase transitions, in which the first non-trivial 
homotopy group of the symmetry breaking group is order 1. Or equivalently, when the vacuum manifold is 
topologically $S^1$ (see for example Ref.~\cite{Sakellariadou:2009ev} for an introduction to Cosmic Strings and
Cosmic Superstrings). As they are topological objects, they cannot end and thus form either loops or infinitely
long strings (i.e.\ they stretch across the horizon). For scales much larger than the thickness of the string
(the `thin string' approximation), the dynamics of Cosmic Strings can be approximated by the Nambu-Goto action,
\be\label{eq:basic_action}
 S = -\mu \int  {\rm d} \tau {\rm d} \sigma \sqrt{ -\gamma}~,
\ee
where $\mu$ is the tension of the string, $(\tau,\sigma)$ are two world sheet coordinates that are respectively
time-like and space-like and $\gamma$ is the determinant of $\gamma_{\alpha\beta}$, the induced world sheet metric given by,
\be
 \gamma_{\alpha\beta} = x_{,\alpha}^\mu x_{,\beta}^\nu g_{\mu\nu}~,
\ee
where $g_{\mu\nu}$ is the space-time metric, and $\alpha, \beta, \dots = \left(1,2\right) =
\left(\tau, \sigma\right)$, whilst $\mu,\nu,\dots = \left(0,1,2,3\right)$ label the space-time coordinates.
 Working in Minkowski space-time (as we will do through out), we can
fix the world sheet gauge freedoms due to the world sheet reparameterization invariance, by making $\gamma_{\alpha\beta}$
conformal. This gives the conditions,
\be\label{eq:gauge_con}
 \dot{x}^2 + x'^2 = 0~, \ \ \ \ \ \ \ \dot{x}\cdot x' = 0~,
\ee
where the dot refers to a derivative with respect to $\tau$ and the dash a derivative with respect to $\sigma$.
Finally we can impose the static gauge conditions, $\tau = t = x^0$, which fixes all the residual freedom.
The equation of motion given by the variation of
 Eq.~(\ref{eq:basic_action}) with respect to $x$, is just the standard wave equation,
which has solutions in terms of arbitrary
right and left movers: ${\bf a}\left(t-\sigma\right)$ and ${\bf b}\left(t+\sigma\right)$,
\be\label{eq:EoM}
 {\bf x} = \frac{1}{2}\left[ {\bf a}\left(t-\sigma\right) + {\bf b}\left(t+\sigma\right) \right]~,
\ee
where $x = \left(t,{\bf x}\right)$.
In terms of these left and right movers, the gauge constraints, Eq.~(\ref{eq:gauge_con}) become especially simple:
$|{\bf a}'| = |{\bf b}'| = 1$, where here the dash refers to the total derivative. Thus the solutions to the
equation of motion can be visualized by considering ${\bf a}'$ and ${\bf b}'$ as curves on the unit sphere.
For the case of closed Cosmic Strings, we have that $\sigma = \sigma + L$ (again working in Minkowski space-time),
where $L$ is the coordinate length of the string. If we take the zero momentum frame i.e.\ the frame moving with
the center of mass of the loop of Cosmic String, then both ${\bf a}'$ and ${\bf b}'$ are also periodic in $L$.\
Moreover, one can show that they satisfy the 'center of mass' conditions,
\be\label{eq:averaging}
 \frac{1}{L}\int_0^L {\bf a}' {\rm d} \sigma = \frac{1}{L}\int_0^L {\bf b}'{\rm d}\sigma = 0~.
\ee

Thus we can visualize the solutions to the equations of motion, that also satisfy the gauge constraints, as
close curves on the unit sphere, that average to zero according to Eqs.~(\ref{eq:averaging}). Because the
two curves ${\bf a}'$ and ${\bf b}'$ are independent, one can easily see that they will generically 
intersect\footnote{the word 'generically' is important here, since it is possible to construct two closed
curves on the unit sphere, with appropriate averages, that do not intersect. However such cases are 
rather special and there is a strong relationship between the two curves. Hence they are not considered
to be generic.}. At intersection, we have ${\bf a}' = {\bf b}'$ and from Eq.~(\ref{eq:EoM}) we see that
such points have velocity given by,
\be
 \dot{\bf x} =  {\bf a}' = {\bf b}'~,
\ee
which, by our gauge constraints, implies that such points have unit velocity. These luminous points are referred
to as cusps. See, for example, Ref.~\cite{Vilenkin_Shellard} for a more detailed discussion of cusps on Cosmic Strings and topological
defects in general. Naturally
one cannot have a massive object moving at the speed of light, instead what this tells us, is that the thin
string approximation breaks down. Despite this, general arguments~\cite{Brandenberger:1986vj} and field theory simulations~\cite{Olum:1998ag}
have shown that cusps still produce vast amounts of both gravitational and particle radiation, making them
candidates for possible observation (see for example Refs.~\cite{Siemens:2006vk,Siemens:2006yp}).

The purpose of this paper is to emphasize what changes (and what remains the same) when we consider Cosmic Superstrings.

\section{Cosmic Superstrings}
Cosmic Superstrings come in two flavors: fundamental F-strings and D-strings (either D$1$-branes or 
D$p$-branes, with $p-1$ transverse directions aligned along the compactified dimensions). They are characterized by
the presence of a world sheet electromagnetic field, $A_{\alpha}$, for which the ends of F-strings are sources~\cite{Johnson}.
This allows Cosmic Superstrings to form three string junctions, in which an F-string ends on a D-string, forming a
bound state FD-string, which carries the conserved flux of $A_{\alpha}$~\cite{Copeland:2003bj}.
The interest in Cosmic
Superstrings comes from the fact that they would be copiously produced at the end of brane inflation~\cite{Barnaby:2004dz}
and hence would be a probe of the early, stringy, universe. They also would naturally have small tensions, which would
help explain the increasingly tight observational upper bounds that have been found for the tensions of one dimensional
defects. In the following we will look at the presence of
cusps on Cosmic Superstrings as a possible way of distinguishing Cosmic Superstrings from their field theoretic
cousins. Of particular interest is the fact that cusps naturally radiate energy into both gravitational waves
and particles. In standard Cosmic Strings, these particles are all possible excitations of the symmetry
group from which the string arose. In Cosmic Superstrings however, all possible excitations available in string
theory might be excited e.g.\ Dilatons~\cite{Damour:1996pv}, Kluza-Klein modes, moduli fields, SUSY particles etc. Thus 
cusps on Cosmic Superstrings are a possible (if remote) direct test of string theory, as well as a consequence
of early universe brane inflation models.

\subsection{A single Cosmic Superstring}\label{sec:single_CSS}
The starting point for this discussion on Cosmic Superstrings is the DBI action\footnote{See, for example, \cite{Johnson}
for a derivation of this action.}, which for a single string is
\be\label{eq:DBI_action}
 S = -\mu \int  {\rm d} \tau {\rm d} \sigma \sqrt{ -|\gamma_{\alpha\beta} + \lambda F_{\alpha\beta}|}~,
\ee
where, $F_{\alpha\beta} = \partial_\alpha A_\beta - \partial_\beta A_\alpha$ is the world sheet electromagnetic
field tensor and $\lambda=2\pi \alpha'$, with $\alpha'$ the Regge slope of String Theory. It can be shown~\cite{Copeland:2007nv}
that the gauge freedoms are again fixed by the constraints given in Eq.~(\ref{eq:gauge_con}) and that these constraints
are compatible with the equations of motion, which are,
\be\label{eq:DBI_EoM}
 \ddot{\bf x} + {\bf x}'' = 0~, \ \ \ \ \dot{P} = P' = 0~,
\ee
where
\be
 P\equiv \frac{\partial {\cal L}}{\partial F_{\tau\sigma}} = \frac{\lambda^2 \mu F_{\tau\sigma}}{\sqrt{-x'^2 \dot{x}^2
- \lambda^2 F_{\tau\sigma}^2}}~,
\ee
is the canonical momenta of the electromagnetic field.

Because the world sheet electromagnetic field is non-dynamical, the dynamics of a Cosmic Superstring are exactly as in the
standard case. The difference arise because of the possibility of a Cosmic Superstring ending, either at a junction or
on a D-brane. One can show~\cite{Davis:2008kg} that for the case of a Cosmic Superstring ending of two parallel D$n$-brane
the left and right movers have the following properties,
\beq
|{\bf a}'| &=& |{\bf b}'| \ \ = \ \ 1~, \nonumber \\
{\bf a}'\left( \xi \right) &=& {\bf a}'\left(\xi\right)~, \nonumber \\
{\bf b}'\left( \xi \right) &=& {\bf b}'\left(\xi\right)~, \nonumber \\
{\bf a}'_{\|}\left(\xi\right) &=& {\bf b}'_{\|}\left(\xi\right)~, \nonumber \\
{\bf a}'_\perp\left(\xi\right) &=& - {\bf b}'_\perp \left(\xi\right)~, \nonumber \\
\frac{1}{2}\int^L_{-L} {\bf a}' \left( \xi\right) {\rm d} \xi &=& \Delta~, \nonumber \\
\frac{1}{2}\int^L_{-L} {\bf b}' \left( \xi\right) {\rm d} \xi &=& -\Delta~,
\eeq
where $\left(\perp, \|\right)$ indicate respectively the directions perpendicular and parallel to the end branes and $\Delta$ is
a vector stretching between the two branes. From the first three of these conditions we can see that, just as in
previous case, we can visualize the solutions as closed curves on the unit sphere. However the fourth and fifth
conditions tell us that the curves giving ${\bf a}'$ and ${\bf b}'$ are no longer independent, instead they are
reflections of each other through a plane parallel to the branes. Finally the sixth and seventh conditions\
show that the 'center of mass' of these closed curves is not zero, but instead depends on the inter-brane
separation.

\begin{figure}
  \includegraphics[height=.3\textheight]{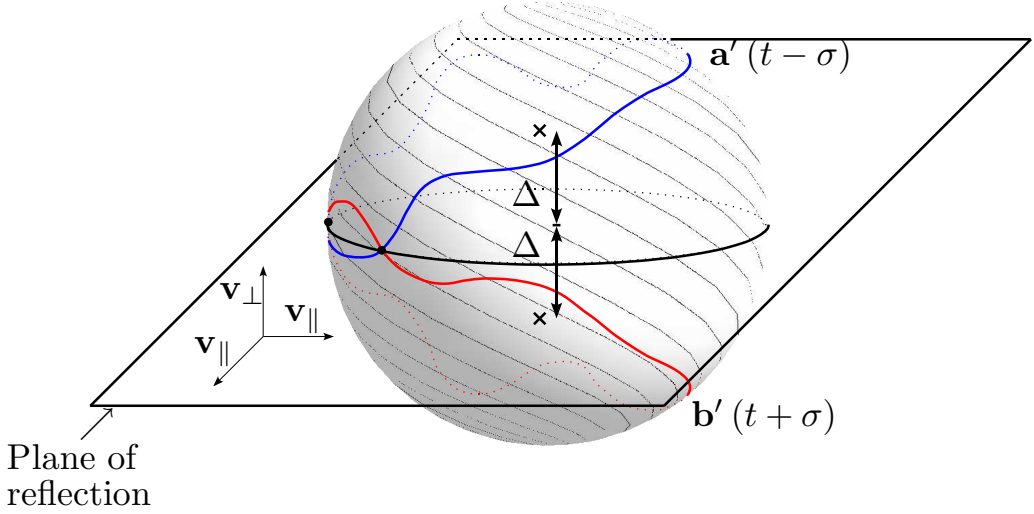}
  \caption{\label{fig:1} The solution for a Cosmic Superstring, governed by the DBI action, ending on two parallel D$2$-branes is given in
terms of its right and left movers: ${\bf a}$ and ${\bf b}$. The boundary conditions imply that their total derivatives
must be closed curves on the unit sphere, that are reflections of each other, with a `center of mass' given by the inter-brane
separation. When these curves intersect, the Cosmic Superstring will have a cusp. Note: ${\bf v}$ is an arbitrary vector that has
been decomposed into components parallel ($v_\|$) and perpendicular ($v_\perp$) to the branes.
}
\end{figure}

The easiest case to consider is a DBI string ending of a pair of D$2$-branes. This situation is depicted in Figure~\ref{fig:1}. It is
clear that if ${\bf a}'$ (or ${\bf b}'$) ever crosses the plane of reflection, then the two curves will intersect and the
Cosmic Superstring will have a cusp. In the limit that the branes are separated by $L$, the coordinate length of the string,
these closed curves become confined to the the points on the sphere furthest from the plane of reflection. Thus they will
not intersect each other. However, when the branes are close together (i.e.\ when $\Delta \approx 0$), we can again see that
intersections (and hence cusps) will be generic. A quantitative analysis of what constitutes `close' in this context is work in
progress, however the result that it is possible to get cusps on non-periodic Cosmic Superstring is robust.

\begin{figure}
  \includegraphics[height=.3\textheight]{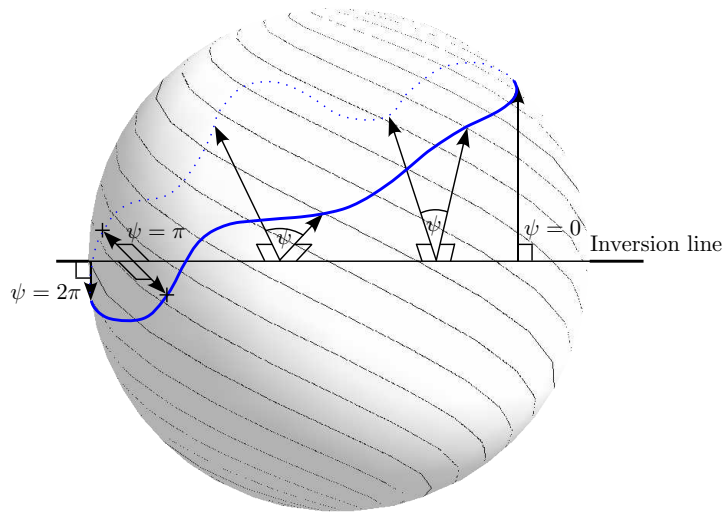}
  \caption{\label{fig:2} For the case of a DBI string ending on a pair of parallel D$1$-branes, one needs to decide if 
a closed curve, reflected through a line, will intersect itself. In the case of the reflection line being encircled
by the closed curve, there will be such an intersection. This can be seen by considering the pair of vectors that are 
perpendicular to the reflection line and end on the closed curve. By continuity there must be at least one point
at which the angle between these vectors $\psi=\pi$ and hence at least one intersection between this closed curve and
its reflection.
}
\end{figure}

The more interesting case for cosmology is that of a DBI string ending on a D$1$-brane, since this is a step towards modeling
a string ending on a three string junction. This can also be visualized as a pair of closed curves on the unit sphere, however
in order to demonstrate that the reflection of a closed curve though a line, will intersect itself, one has to consider some,
essentially, topological arguments. In particular, consider the two vectors that are perpendicular to the reflection line and
end on the closed curve. If the reflection line is encircled by the curve, then the angle between these two vectors is $0$ at
one extreme and $2\pi$ at the other (see Figure~\ref{fig:2}). By continuity, the must be a point at which the angle between
the vectors in $\pi$ and thus at least one pair of points which are mapped into each other by the reflection.

It fact it can be shown~\cite{Davis:2008kg}, that in approximately half of all possible cases such intersections will occur,
at least whilst the end branes are close together. Thus in a significant proportion of the time, one would expect 
to have cusps on a Cosmic Superstring ending on a pair of D$1$-strings.

\subsection{Three string junctions}
In the previous section it was shown that it is possible to have cusps on Cosmic Superstrings that end on D-branes.
Here we want to consider the case of a three string junction, formed after the collision between an F- and a
D-string, which combine to produce a bound state FD-string~\cite{Copeland:2006eh}. The DBI action for such a 
junction, again working in the conformal gauge, is  given by~\cite{Copeland:2007nv,Copeland:2006if},
\beq\label{eq:3stringaction} 
S &=& - \sum_i \mu_i \int{\rm d}\tau\int_{0}^{L_{i}(\tau)}{\rm
  d}\sigma \sqrt{ -{x'_i}^2 {\dot{x}_i}^2 -\lambda
  (F^i_{\tau\sigma})^2 } \nonumber \\
&&\hspace{-1.8truecm} + \sum_i \int{\rm
  d}\tau \left\{{\bf f}_i(\tau) \cdot [
  x_i(\tau,L_i(\tau))-\bar{x}(\tau)]
+g_i\left[A^i_{\tau}(\tau,L_i(\tau))+ \dot{L}_i A^i_\sigma
  (\tau,L_i(\tau)) - \bar{A}(\tau) \right]\right\}~,
\eeq 
where $\mu_i$ is the tension of the $i^{\mathrm{th}}$-string, and
$A^{i}_{\alpha}$ and $F^{i}_{\alpha\beta}$ are the gauge field and
gauge field strengths respectively on the
$i^{\mathrm{th}}$-string. The Lagrange multipliers ${\bf f}_i(\tau)$
constrain the strings to meet at the junction (with world line $\bar{x}(\tau)$), whilst
the Lagrange multipliers $g_i(\tau)$ impose to the component of each
string's gauge field that is tangential to the junction world-line to
coincide with the junction gauge field $\bar{A}(\tau)$. Note that, just as for cusps,
the presence of a junction breaks the thin string limit (still present in the DBI-action),
however numerical calculations show that the above action is a good approximation to the
dynamics of such systems~\cite{Salmi:2007ah}.

The equations of motion for each string in this action is the same as that of a single
DBI-string (Eqs.~(\ref{eq:DBI_EoM})), however the boundary condition at the 
junction end (which we fix to be at the $\sigma_i = L_i(t)$ end of the strings)
is given by the variation of Eq.~(\ref{eq:3stringaction}) with respect to the
Lagrange multipliers and with respect to $\bar{x}(\tau)$ and
$\bar{A}(\tau)$. This gives the following set of conditions:
\be\label{eq:juncbc2}
x_i\left(\tau,L_i(\tau)\right)=\bar{x}\left(\tau\right)
\ee
and
\be
A^i_{\tau}\left(\tau,L_i(\tau)\right)+ \dot{L}_i A^i_\sigma
   \left(\tau,L_i(\tau)\right) = \bar{A}(\tau)~.
\ee
\be\label{eq:juncbc1}
\sum_i\bar{\mu}_i\left({x_i^{\mu}}'+\dot{L}_{i}\dot{x}_i^{\mu}\right)=0
\ee
and
\be\label{eq:fstringcons}
\sum_{i}p_{i}=0\ ,
\ee
where the effective string tension $\bar{\mu}_i$ is a combination of the tension and 
the electromagnetic field strength on the $i^{\rm th}$ string. Note that as in the previous 
cases it is possible to fix the temporal gauge condition ($\tau = x^0 = t$) and that
we can no longer fix the coordinate length of each string to be a constant, instead
we have $L_i\left(\tau\right)$.

These equations can be used, along with the equations of motion, to find~\cite{Copeland:2006eh,Davis:2008kg}
\be
\label{eq:sols}
\frac{\bar{\mu}_1}{\bar{\mu}_1+\bar{\mu}_2+\bar{\mu}_3} \left(
1-\dot{L}_i\right)= \frac{ M_1 \left( 1-c_{23}\right)}{M_1\left(
  1-c_{23}\right)+ M_2\left( 1-c_{13}\right)+ M_3\left(
  1-c_{12}\right)}~,
\ee
with cyclic permutations giving expressions for $\dot{L}_2$ and
$\dot{L}_3$, where
\be
M_1 = \bar{\mu}_1^2 - \left(\bar{\mu}_2 - \bar{\mu}_3 \right)^2~,
\ee
with cyclic permutations giving $M_2$ and $M_3$ and 
scalar products are abbreviated to
\be
c_{ij}={\bf a}'_i(t-L_i(t))\cdot{\bf a}'_j(t-L_j(t))~.
\ee

In order to make progress with this, we restrict our attention to the specific case of an F-, D- and FD-string junction 
(until now, the equations are applicable to any three DBI strings). For this situation we have the following relationship
between the effective tensions $\bar{\mu}_i$:
\be
\bar{\mu}_1 = 1~, \ \ \ \bar{\mu}_2 = \frac{1}{g_{\rm s}}~,
\ \ \ \bar{\mu}_3 = \sqrt{ 1+\frac{1}{g_{\rm s}^2} } = \frac{1}{g_{\rm s}} +
\frac{g_{\rm s}}{2} + {\cal O}\left( g_{\rm s}^3\right)~,
\ee
where $g_{\rm s}$ is the perturbative string coupling and we have taken the $1^{\rm st}$ string to be the 
F-string, the $2^{\rm nd}$ string to be the D-string and the $3^{\rm rd}$ string to be the bound state FD-string.
To leading order in $g_{\rm s}$, 
Eqs.~(\ref{eq:sols}) give,
\beq \label{eq:b1pert}
\left(S_{23}- 2S_{13}-2S_{12}\right)
     {{\bf b}'_1} &=& S_{23} {{\bf a}'_1} -2S_{13}{{\bf a}'_2} -2S_{12}{{\bf a}'_3} ~,\\
\label{eq:b2a3}
{{\bf b}'_2} &=& {{\bf a}'_3} ~,\\
\label{eq:b3a2}
{{\bf b}'_3} &=& {{\bf a}'_2} ~,
\eeq
and
\be\label{eq:L1dotpert}
\dot{L}_1=1-  \frac{S_{23}}{S_{12}+S_{13}}
  ~,\quad
\dot{L}_2= \frac{S_{12}-S_{13}}{S_{12}+S_{13}}
  ~,\quad
\dot{L}_3=  \frac{S_{13}-S_{12}}{S_{12}+S_{13}}
  ~,
\ee
where $S_{ij}=\frac{1}{2}(1-c_{ij})$ and the notation
\be 
{\bf a}'_i = {\bf a}'_i(t-L_i(t))\quad\mathrm{and}\quad
{\bf b}'_i = {\bf b}'_i(t+L_i(t))~,
\ee
has been used.

The physical interpretation of this is clear, Eqs.~(\ref{eq:b2a3}) and (\ref{eq:b3a2}) show that the in-movers
on (heavy) D-string become the out-movers on the (heavy) bound state FD-string and vice verse. Also,
we have that $\dot{L}_2 + \dot{L}_3 = 0$. Thus the D- and the FD-strings are essentially behaving as a
single long string, entirely unaffected (to order $g_{\rm s}$) by the presence of the (light) F-string
or the junction. Whilst the (light) F-string moves in the background of the heavy string, it does so without
back-reacting on its motion, similar in many ways to the Born-Oppenheimer approximation in condense matter physics.

For the F-string the junction has exactly the same boundary conditions as a standard D$1$-brane (albeit a moving
D$1$-brane), which we have already seen in Section~(\ref{sec:single_CSS}). There we considered only static branes 
as boundary conditions, however the extension to moving branes can be done. Indeed, because the presence of
cusps is a topological feature, a general class of deformations to the static case can be considered and shown
to leave the presence of cusps unaltered. In particular the additional rotation that follows the reflection 
considered in Section~(\ref{sec:single_CSS}), that is implied by Eq.~(\ref{eq:b1pert}) belongs to such a class
and hence does not affect the presence of cusps.

\section{Conclusions}
We have shown that for a three string junction formed from F-, D- and FD-strings, the heavy D- and FD-strings
behave as a single string, whilst the F-string moves as though the junction were a standard Dirichlet 
boundary conditions. This has several important consequences for the evolution of Cosmic Superstrings.

Firstly, the fact that the collision of F- and D-strings and the formation of bound states does not alter (to order
$g_{\rm s}$) the dynamics of the heavy D-strings, implies that scaling will still be reached in this sector of
the theory. There has been some numerical evidence~\cite{Urrestilla:2007yw} that suggests that this is indeed the case
(at least for the case of the non-Albelian model of Cosmic Superstrings). However it remains to been seen how
accurate the $g_{\rm s}\rightarrow 0$ limit is for such a model, particularly on the motion of loops of the
heavy string (which can and will still form). It is possible that as the number of junctions increase, and the 
number and type of bound states increase (FFD-, FFFD-strings etc.), the probability of forming a `free' 
heavy loop drops to zero. In which case the subsequent motion of the loops would be correlated with that of the
long (heavy) strings.

Secondly, during collisions between F- and D-strings, the F-string will break onto the D-string, forming a pair of
junctions. For any F-string ending of two such junctions, we have demonstrated that it will contain cusps. This
represents an entirely new energy loss mechanism for the system, which may have significant consequences for the
final scaling ratios (if scaling is indeed reached) of the energy in each sector. In addition, these cusps
represent a way of distinguishing Cosmic Superstrings from standard Cosmic Strings, if they can be discriminated
from the cusps present on loops of string. As mentioned in Section~(\ref{sec:intro}), the radiation produced by
such cusps could, in principle, excite all possible String theory modes, making them a possible direct probe
of string theory as well as Cosmic Superstrings.

Whilst we have shown that cusps are likely to be important features of strings ending on junctions,
a quantitative analysis of the relative importance of loops and junctions, in the formation of cusps is still needed. In
addition, the issue of small scale structure and its effect of the presence of cusps needs to clarified. In fact
in standard Cosmic Strings, small scale kinks are expected to form at every collision event, however in the case
of Cosmic Superstrings, many of these collisions will form junctions. This would seem to suggest that cusps 
are even more likely on Cosmic Superstrings, compared to standard field theoretic strings, however one would need
to confirm this qualitative expectation with rigorous numerical or analytical work.

%%%%%%%%%%%%%%%%%%%%%%%%%%%%%%%%%%%%%%%%%%%%%%%%
%% BACKMATTER
%%%%%%%%%%%%%%%%%%%%%%%%%%%%%%%%%%%%%%%%%%%%%%%%

\begin{theacknowledgments}
This article is based on the longer and more detailed paper written with
Anne-Christine Davis, Senthooran Rajamanoharan, and Mairi Sakellariadou~\cite{Davis:2008kg}.

This work was supported in part by the NSF grant PHY0854743, The George A.
and Margaret M. Downsbrough Endowment and the Eberly research funds
of Penn State.
\end{theacknowledgments}



\bibliographystyle{aipproc}   % if natbib is available

\bibliography{Nelson}

\IfFileExists{\jobname.bbl}{}
 {\typeout{}
  \typeout{******************************************}
  \typeout{** Please run "bibtex \jobname" to optain}
  \typeout{** the bibliography and then re-run LaTeX}
  \typeout{** twice to fix the references!}
  \typeout{******************************************}
  \typeout{}
 }

\end{document}